 \documentclass[sigconf]{acmart}
\usepackage{framed}
\usepackage{morewrites}
\usepackage{float}
\usepackage{pifont}
\usepackage{wasysym}
\usepackage[most]{tcolorbox}
\usepackage{lipsum} 
\usepackage{xcolor}
\usepackage{enumitem}

\newcommand{\revise}[1]{{\color{black} #1}}

\AtBeginDocument{%
  }

\setcopyright{acmlicensed}
\copyrightyear{2018}
\acmYear{2018}
\acmDOI{XXXXXXX.XXXXXXX}
\acmConference[Conference acronym 'XX]{Make sure to enter the correct
  conference title from your rights confirmation email}{June 03--05,
  2018}{Woodstock, NY}
\acmISBN{978-1-4503-XXXX-X/2018/06}

\begin{document}

\title{“Before, I Asked My Mom, Now I Ask ChatGPT”: 
\\Visual Privacy Management with Generative AI for Blind and Low-Vision People}

\author{Tanusree Sharma}
\authornotemark[1]
\email{tanusree.sharma@psu.edu}
\orcid{0000-0003-1523-163X}
\affiliation{%
  \institution{Pennsylvania State University}
  \city{State College}
  \state{PA}
  \country{USA}
  \postcode{16802}
}

\author{Yu-Yun Tseng}
\orcid{0000-0002-4223-0483}
\affiliation{%
 \institution{Computer Science, University of Colorado}
 \city{Boulder}
 \state{Colorado}
 \country{USA}}
\email{yu-yun.tseng@colorado.edu}

\author{Lotus Zhang}
\orcid{0000-0002-6315-9970}
\affiliation{%
  \institution{Human Centered Design and Engineering, University of Washington}
  \city{Seattle}
  \country{USA}}
\email{hanziz@uw.edu}

\author{Ayae Ide}
\affiliation{%
  \institution{Pennsylvania State University}
  \city{State College}
  \state{PA}
  \country{USA}
\email{ayaeide@psu.edu}
}

\author{Kelly Avery Mack}
\affiliation{%
  \institution{Human Centered Design and Engineering, University of Washington}
 \city{Seattle}
  \country{USA}
\email{kmack3@uw.edu}
}
\author{Leah Findlater}
\orcid{0000-0002-5619-4452}
\affiliation{%
  \institution{Human Centered Design and Engineering, University of Washington}
  \city{Seattle}
  \country{USA}}
\email{leahkf@uw.edu}

\author{Danna Gurari}
\orcid{0000-0003-1306-0283}
\affiliation{%
 \institution{Computer Science, University of Colorado}
 \city{Boulder}
 \state{Colorado}
 \country{USA}}
\email{danna.gurari@colorado.edu}

\author{Yang Wang}
\orcid{0000-0003-3641-8241}
\affiliation{%
  \institution{Information Sciences, University of Illinois at Urbana-Champaign}
  \city{Champaign}
  \country{USA}}
\email{yvw@illinois.edu}

\renewcommand{\shortauthors}{Sharma, et al.}
\renewcommand{\shortauthors}{Sharma et al.}
\newcommand{\yang}[1]{{\color{blue}{[Yang: #1]}}}

\begin{abstract}
Blind and low vision (BLV) individuals use Generative AI (GenAI) tools to interpret and manage visual content in their daily lives. While such tools can enhance the accessibility of visual content and so enable greater user independence, they also introduce complex challenges around visual privacy. In this paper, we investigate the current practices and future design preferences of blind and low vision individuals through an interview study with 21 participants. Our findings reveal a range of current practices with GenAI that balance privacy, efficiency, and emotional agency, with users accounting for privacy risks across six key scenarios: self-presentation, indoor/outdoor spatial privacy, social sharing, and handling professional content. Our findings reveal design preferences, including on-device processing, zero-retention guarantees, sensitive content redaction, privacy-aware appearance indicators, and multimodal tactile mirrored interaction methods. We conclude with actionable design recommendations to support user-centered visual privacy through GenAI, expanding the notion of privacy 
and responsible handling of others’ data.


\end{abstract}



\maketitle

\section{Introduction}

Generative AI (GenAI) has increasingly become integrated into everyday life for blind and low vision (BLV) people~\cite{adnin2024look, bendel2024can, raman2024exploring, flores2025impact, msr,letsenvisionEnvisionPerceive, seeingaiSeeingTalking}, supporting a wide range of tasks.  For example, there has been recent growth in the number of available GenAI visual interpretation tools, such as those offered by Be My Eyes~\cite{bemyeyesHome}, Envision AI~\cite{letsenvisionEnvisionPerceive}, and Seeing AI~\cite{seeingaiSeeingTalking}, as well as image generation tools, such as GenAssist~\cite{huh2023genassist}, AltCanvas~\cite{lee2024altcanvas}.  These tools are used for tasks such as recognizing objects, describing scenes, answering visual questions~\cite{xie2025beyond, penuela2025towards}, and supporting spatial navigation 
~\cite{tokmurziyev2025llm, hao2024chatmap}.


While these advances introduce new opportunities, they also raise concerns about visual privacy when BLV individuals leverage GenAI tools to manage, consume, and share visual content. Prior research has explored visual privacy risks in both intentionally and unintentionally sharing private content~\cite{Bennett18, Gonzalez22, Park20, Harada13, Adams13, Alharbi2022, Zhang2023ImageAlly, stangl2023dump, Brunton2015, tseng2024biv}, human assistance versus visual assistive technologies (VATs) support~\cite{Alharbi2022, stangl2023dump}, as well as in context of concern BLV users have for sharing and managing contents of their own and others with VATs~\cite{akter2020uncomfortable, stangl2020visual, ahmed2016addressing}. With GenAI introducing new ways of interacting with visual information, it is important to understand not only how existing visual privacy concerns may manifest differently or evolve with these emerging technologies, but also how GenAI might actively support BLV users in safeguarding visual privacy.\footnote{Based on prior work, visual privacy refers to the safeguarding and management of sensitive visual information that could be shared or disclosed in everyday life~\cite{stangl2023dump, ahmed2015privacy}} 

Despite significant commercial and public attention towards the successes of GenAI for blind and low vision people~\cite{letsenvisionEnvisionPerceive, seeingaiSeeingTalking, bemyeyesHome}, we know little about how blind people use GenAI tools in managing their privacy~\cite{gurari2019, ahmed2015privacy, ahmed2016addressing, Voykinska16, zhang2024designing}. Specifically, it is unclear how generative AI may be supplementing or replacing tasks traditionally supported by human assistants or VATs, and how users conceptualize and evaluate these tools in relation to their privacy concerns.
Filling this gap, we explore two research questions: (\textbf{RQ1:}) How do blind and low vision people currently use Generative AI tools to manage visual privacy? and (\textbf{RQ2:}) What design opportunities do blind and low vision people envision in future Generative AI tools to support visual privacy management?  

To this end, we conducted semi-structured interviews with 21 individuals from the United States who are blind and have low vision to understand their current practices and design expectations for GenAI to manage visual privacy. Acknowledging the context-dependent nature of visual privacy management, through a scenario-driven inquiry, we grounded our interviews in six different common scenarios\footnote{Six key use cases in which visual privacy can emerge: (1) Self Appearance and Impression Management, (2) Indoor Spatial Privacy, (3) Sharing Photos on Social Media, (4) Visual Privacy with Employers when Sharing Content, (5) Visual Privacy as BLV Professionals when Assessing Others’ Content, and (6) Outdoor Spatial Privacy.} where privacy concerns can occur. We investigate to what extent study participants manage privacy with GenAI tools in these scenarios and explore expectations for how GenAI \emph{should} be designed to better support visual privacy. 

In summary, the main contributions of this work are as follows:

(1) The first in-depth empirical analysis of how BLV individuals use GenAI tools to manage visual privacy. Our findings point to an emerging, yet complex, use of GenAI to retain emotional and informational privacy in sensitive situations, such as interpreting pregnancy tests and mammograms, despite limitations in accuracy. Participants often emphasized values of independence and personal privacy as key factors influencing their decision to choose GenAI over VATs or human assistance.

(2) Our findings highlight the emerging use of GenAI for visual privacy management, shaped by participants’ current experiences, perceived limitations, and expectations for future design. For instance, using GenAI for self appearance is not limited to how the individual looks, but also to managing privacy involving bodily presentation or vulnerability
to manage emotional agency. Our result also indicates privacy as institutional responsibilities where BLV professionals manage others’ content. These expectations of BLV users reflect how they negotiate trade-offs between autonomy, risk, and convenience, taking into account emotional vulnerability and the privacy of others.

(3) We propose a set of privacy-preserving design interventions that are grounded in participants’ lived experiences, such as on-device processing, a compliance-aware secure toolkit for BLV professionals, and personalized appearance indicators.
\section{Related Work}
Our research is informed by prior literature on visual privacy for people with visual impairment, privacy-preservation technology, and the current technological shift in accessibility with GenAI.

\vspace{-2mm}
\subsection{Use of Generative AI for Visual Accessibility}
Recently, GenAI tools (e.g., ChatGPT, Google Gemini, Microsoft Copilot, and Claude) have been incorporated into a wide variety of domains, including for education~\cite{hazzan2024generative, kazemitabaar2023studying}, communication~\cite{chen2023closer}, content creation~\cite{inie2023designing}, and visual interpretation~\cite{shu2023audio}. While much of the progress was initially for text-based applications, we are increasingly seeing a move into GenAI for visual content.\footnote{Industry successes follow strong foundational research. A notable example is CLIP has shown impressive zero-shot performance~\cite{radford2021learning}
in downstream applications, ranging from object detection to 3D applications~\cite{bangalath2022bridging, liang2023open, rozenberszki2022language, ni2022expanding}, and has been adapted for video applications~\cite{ni2022expanding, wang2021actionclip, rasheed2023fine}.
More recently, multimodal integration has advanced with models like Flamingo~\cite{alayrac2022flamingo}, BLIP-2~\cite{li2023blip} 
MiniGPT-4~\cite{zhu2023minigpt}, and LLaVA ~\cite{liu2024visual} leveraging web-scale image-text data for improved multimodal chat capabilities. Some works extend LLMs for video comprehension~\cite{maaz2023video, radford2021learning, chiang2023vicuna, li2023videochat, liu2024visual}, introducing Video-ChatGPT, a model combining a video-optimizer for enhanced understanding.} Blind people are engaging with the growing ecosystem of such generative AI tools, including via accessibility tools such as Be My Eyes and Envision AI~\cite{thevergeEyesOffers, forbesEnvisionAdds}. Recent work on GenAI for blind users has explored accessible image generation, alt text, scene descriptions, and verification strategies, and has emphasized the need for explainable, customizable, and context-aware GenAI tools to support access, creativity, and trust ~\cite{das2024provenance, gonzalez2024investigating, chang2024worldscribe, huh2023genassist}.
However, GenAI tools are rarely designed with the unique privacy needs of BLV users in mind~\cite{adnin2024look}, which could potentially result in both functional gaps and potential harms. Filling this gap, we explore how blind individuals are currently using generative AI tools for privacy management and the purposes these tools serve in their daily lives. 

\vspace{-2mm}
\subsection{Visual Privacy Management with GenAI}
Much of the literature on privacy management in GenAI focuses on sighted users~\cite{sun2024empowering, chen2025clear, wang2024lave, zhou2024rescriber}. Recent studies have proposed interventions to mitigate privacy risks, yet these efforts primarily address sighted users' need, for instance, Chong et al.~\cite{chong2024casper} developed a system for prompt sanitization using web-based LLMs to tackle excessive disclosure. 
A recurring theme across prior work is the inherent trade-off between privacy, utility, and convenience, often framed as a core challenge in GenAI. Ma et al. ~\cite{ma2025raising} introduced an LLM to raise awareness of location-based privacy risks in images by detecting subtle visual cues. Similarly, CLEAR~\cite{chen2025clear} provides real-time contextual risk feedback during sensitive data entry in tools like ChatGPT and Gmail to facilitate just-in-time privacy literacy. Zhou~\cite{zhou2024rescriber} proposed Rescriber, which abstracts sensitive content before the prompt to reduce data exposure. Complementary tools such as Adanonymizer and PrivacyAsst~\cite{zhang2024privacyasst, zhang2024adanonymizer} offer user-facing privacy controls for adjusting privacy-performance settings with cryptographic techniques to protect sensitive visual media. 
Despite these advancements, current design approaches remain limited in addressing the privacy needs of BLV users, a gap that underscores the need for inclusive GenAI privacy solutions.


\vspace{-2mm}
\subsection{BLV Individuals' Visual Privacy Concerns and Goals}

Some research revealed the visual privacy management interests of BLV individuals~\cite{Bennett18, Gonzalez22, Park20, Harada13, Adams13, ahmed2015privacy, ahmed2016addressing, Voykinska16, kaushik2023guardlens}. BLV users share images and videos both to socially connect~\cite{Bennett18, Gonzalez22, Park20} and to receive assistance in understanding their surroundings. Privacy-sensitive content includes financial and medical documents (e.g., prescriptions, pregnancy tests), identifiable personal information (e.g., addresses, faces), and images that may affect how others view them (i.e., impression management), including unflattering or awkward photographs, disorganized living spaces~\cite{stangl2020visual, akter2020uncomfortable, sharma2023disability}. Additional concerns arise when bystanders’ privacy is at risk or when disclosures, intentional or accidental, may impact social relationships~\cite{stangl2020visual, akter2020uncomfortable}.

Pioneering visual interpretation technologies centered on \emph{human assistance}, including from employees (e.g., Aira~\cite{aira_aira_2020}), volunteers (e.g., Be My Eyes~\cite{BeMyEyes}), or paid crowdworkers (e.g., VizWiz~\cite{bigham_vizwiz_2010}), while subsequent research showed how to reduce or eliminate human effort through automation~\cite{wu2017automatic, wu2016ask, guinness2018caption,lee2022opportunities, xie2022iterative}, eventually culminating in today's widely used industry products, including Be My Eyes' Be My AI app~\cite{bemyeyesHome}, Envision AI's glasses-based technology~\cite{letsenvisionEnvisionPerceive}, and Microsoft's Seeing AI app~\cite{seeingaiSeeingTalking}. While prior research on visual privacy for blind people has primarily focused on visual interpretation services~\cite{stangl2020visual, akter2020uncomfortable}, blind users’ strategies for managing privacy extend beyond this domain. Their practices intersect with broader themes in privacy research with GenAI, and contextual privacy decision-making~\cite{nissenbaum2004privacy, acquisti2005privacy}. In our work, we investigate how they their expectation and need can shape the design of privacy-aware GenAI. 


\section{Method}
\label{Methods} 
To investigate blind and low-vision people's current use of GenAI tools for visual privacy and identify future design opportunities, we conducted a semi-structured interview study. Below, we describe our interview protocol, participant recruitment, data collection, and data analysis.   

\begin{table*}[h!]
\small
\centering
\caption{Visual Privacy Scenarios for Contextual Investigation to Understand Limitations \& Design Opportunities of GenAI}
\label{table:scenario}
\begin{tabular}{p{5cm}p{8cm}p{1.5cm}}
\hline
\textbf{Scenario} & & \textbf{Type} \\
\hline
Privacy Scenario 1: Self Appearance: Impression Management & It refers to conscious effort to control how one’s physical appearance is perceived by others, especially in social or professional settings &Own Privacy \\
Privacy Scenario 2: Indoor Spatial Privacy & A person's ability to control and access their space, both physical and perceived within a specific indoor environment &Own Privacy\\
Privacy Scenario 3: Sharing Visual Content in Social Media & Process of reviewing what private details are visible in visual content before sharing them in social media & Own \& Others' Privacy\\
Privacy Scenario 4: Visual Content Privacy when Sharing with Others & Review and control of sensitive information, such as personal, financial, when sharing with others (e.g. employer)& Own Privacy\\
Privacy Scenario 5: Visual Privacy as BLV Professionals & Accessing and reading private visual information like documents of others, while keeping personal details secure and confidential as & Others' Privacy \\
Privacy Scenario 6:  Outdoor Spatial Privacy  & A person's ability to control their personal space, and freedom from intrusion when in public or outdoor settings, such as airports, stations, streets, etc. &  Own \& Others' \\
\hline
\end{tabular}

%
\label{tab:scenarios}
\end{table*}

\subsection{Interview Protocol}\label{Methods_protocol}

The semi-structured interview protocol was approved by the Institutional Review Board (IRB) at Penn State, and can be found in Appendix~\ref{appendix}. The protocol included two sections: (a) current use of GenAI with an emphasis on visual privacy, and (b) perspectives on design improvements of GenAI in common privacy-related scenarios.

In the first section, we investigate participants’ everyday use of GenAI, with a focus on whether and how these tools are used for tasks related to visual privacy. 
Based on participants' response, if they did not mention privacy-related uses of GenAI tools, we followed up with the question~\textit{``Have you used any of these tools you mentioned earlier for a purpose related to privacy?''} 
To facilitate a shared understanding during discussions on design improvements, we provided a verbal explanation of~\textit{``visual privacy~\footnote{Visual Privacy for Blind and low vision users refers to the safeguarding and management of sensitive visual information that could be shared or disclosed through the use of Generative AI tools. This includes but is not limited to the protection of content such as medical records, financial information, or any other visual data that might be considered private when engaging in daily activities. For instance, when using these tools to receive descriptions of potentially sensitive content, whether it’s medical or financial records or when scanning surroundings for navigation}''} to ensure a shared understanding. 

In the second section, we investigated current design opportunities of GenAI by examining how it is currently used, along with participants' perceived challenges and limitations of GenAI. To guide this investigation, we introduced a set of context-specific scenarios where privacy concerns may occur, inspired by prior work~\cite{stangl2020person, gurari2018vizwiz, sharma2023disability, li2022feels, voykinska2016blind, ahmed2015privacy, feng2024understanding, bandukda2020places}. 
(i) self-appearance and impression management (~\cite{sharma2023disability, brady2013visual, li2022feels}), (ii) managing indoor spatial privacy in shared environments (~\cite{faccanha2020m, xie2024bubblecam}), (iii) sharing visual content on social media (~\cite{voykinska2016blind, morris2016most, ahmed2015privacy}), (iv) privacy in document sharing (~\cite{akter2020uncomfortable, feng2024understanding, stangl2023dump}), (v) visual privacy when managing content for others~\cite{hayes2019cooperative, akter2022shared, wahidin2018challenges}, and (vi) outdoor spatial privacy~\cite{bandukda2020places, jonnalagedda2014enhancing, ahmed2015privacy}.  
\revise{We chose these scenarios based on (a) relevancy of the scenario to visual privacy with GenAI ; (b) gap identified in prior work in visual privacy; and (c) diversity of context representing BLV users. Our goal is to fill the gap in exploring the limitations and design opportunities for these use-cases for GenAI.}
We designed a set of questions to (i) identify whether/ how participants currently use or would consider using GenAI for privacy management in these scenarios (see Table~\ref{tab:scenarios}), (ii) uncover points of friction where GenAI supports or challenges established privacy expectations, and (iii) gather participants’ reasoning about potential design improvements within their privacy needs. 

\subsection{Participant Recruitment and Demographic Information}

We initially conducted a screening survey to identify participants who satisfied our inclusion criteria: participants had to (1) use Generative AI tools, and (2) be 18+ years old. We leveraged mailing lists through organizations serving people who are blind~\cite{national_federation_of_the_blind_blind_nodate}.
We moved forward with 21 people for our interview study. They represented a broad range of demographics. As shown in Table~\ref{table:demographics}, participants were 25-54 years old, with 13 identifying as female and eight as male, with open-ended descriptions for gender. Participants' completed levels of education varied, with eight having completed high school, seven having obtained a Bachelor's degree, five having obtained a Master's degree, and one having obtained a Ph.D. Fourteen of them were identified as being totally blind, while seven had some light perception. Participants mostly commonly reported using the following GenAI tools: ChatGPT, Gemini, Perplexity, Be My AI, Envision AI, vdScan, Microsoft Copilot, Google NotebookLM, Replika, and PiccyBot.

\subsection{Data Collection \& Analysis}

For data collection, the first author conducted the interviews with the 21 participants remotely over Zoom between November 2024 and January 2025. The interview duration was an hour each. 
For data analysis, the first author led a thematic analysis~\cite{fereday2006demonstrating, braun2012thematic} of the transcriptions alongside another with a research assistant. Two researchers independently read 20\% of the interview transcripts, developed codes, and compared them until arriving at a consistent codebook. Once the codebook is ready, two researchers applied the codebook to 10\% new data with inter-coder reliability calculated as 0.87 for Cohen's Kappa (a good score as noted in \cite{fleiss2013statistical}).  After finalizing the codebook, the researchers divided the remaining data between them for coding and spot-checked all coded transcripts with each other to ensure ongoing consistency in coding. 
We followed an open coding and deductive analysis method to explore participants’  practices, challenges, and design suggestions of GenAI for visual privacy. We group 102 low-level codes into sub-themes, and further extracted main themes. We iterated this process to finally produce approximately 29 themes to interpret the results. 


\begin{table*}[h]
\small
\centering
\caption{Participant demographics and background.}
\begin{tabular}{l l l l l l l }
\hline
\textit{Participant ID} & \textit{Gender} & \textit{Age} & \textit{Country} & \textit{Blindness}   & \textit{Current GenAIs Used}  \\
\hline
P1 & Male & 25-34 & the US & Totally Blind &  ChatGPT, Gemini\\
P2 & Female & 35-44 & the US & Some Light Perception&  SeeingAI, Be my eyes, AIRA explorer, Microsoft Designer \\
P3 & Female & 35-44  & the US & Totally Blind  & ChatGPT \\
P4 & Female & 35-44 & the US & Totally Blind  & ChatGPT, Gemini \\
P5 & Female & 25-34 & the US & Totally Blind  &  Be My Eyes, Envision AI\\
P6 & Male & 25-34 & the US & Totally Blind  & ChatGPT, Envision, Be My Eyes, and Seeing AI \\
P7 & Male & 35-44 & the US & Totally Blind  &  Seeing AI, Microsoft Copilot, Be My AI\\
P8 & Female & 45-54 & the US & Some Light Perception & Be My Eyes \\
P9 & Male & 35-44 & The US & Totally Blind  & Gemini, ChatGPT\\
P10 & Female & 35-44  & the US & Totally Blind  &  Be My Eyes\\
P11 & Female & 35-44  & the US &Totally Blind  & ChatGPT\\
P12 & Male & 18-24 & the US & Totally Blind  & ChatGPT, Be My Eyes, Google LM notebook, Gemini\\
P13 & Female & 35-44 & the US & Some Light Perception& Seeing AI, ChatGPT, Be My AI, Envision, CoPilot \\
P14 & Male & 25-34 & the US & Some Light Perception  & Be My AI, Aira AI \\
P15 & Female & 45-54 & the US & Some Light Perception  & Seeing AI, ChatGPT, Be My AI \\
P16 & Female & 25-34 & the US & Some Light Perception & Perplexity, ChatGPT, Meta AI, replika, piccybot aira ai  \\
P17 & Female & 25-34 & the US  & Some Light Perception & ChatGPT, Seeing AI\\
P18 & Female & 35-44 & the US & Totally Blind & ChatGPT \\
P19 & Female & 45-54 &  the US & Totally Blind   &gemini, meta ai, replika be my ai piccybot aira ai \\
P20 & Male & 25-34 &  the US& Totally Blind  & Perplexity, chat gpt, claude\\
P21 & Male &35-44 &  the US& Totally Blind  & Seeing AI, chat gpt\\
\hline
\end{tabular}%
\label{table:demographics}
\end{table*}

\begin{table*}
  \small
    \begin{tabular}{p{2cm}p{8.75cm} p{1.5cm} p{1.5cm}}
    \toprule
        {\bf Content} & {\bf Representative Quote} & {\bf Benefit} & {\bf Alternatives} \\
    \midrule
  \textbf{Medical }& & \\ 
  Pregnancy test & \textit{``I read pregnancy test. I didn’t want to rely on someone else to help with something so personal. That’s where I really think about privacy. I used Be My AI if it could read those results to me clearly and privately. (P1)} & Privacy& Family/Friend\\
 Mammogram &	\textit{``I privately understand the report without asking others (P17)''} & Privacy& Aira agents\\ 
 Prescription labels &	 \textit{``I’ve also used regular Be My Eyes to get help reading prescription labels and directions, and I’ve used Seeing AI for that too. Quite accurate (P13)''}&	Convenience &  Aira\\
 Medicine bottle	&\textit{``I navigate stores to find a medicine bottle with be my AI and envision. In one case, repeatedly photographing a medicine bottle led to suspicion from a store employee.''} (P17) &	Convenience and Independence & Employee \\
 \hline
 \textbf{Finance}\\ 
 Bank statement&\textit{``I scanned the bank statement. I do wonder if the GenAI is saving that information, my account number or address.''} (P12) &	Convenience & Aira, Family\\
 
Tax document&	\textit{``at night—I wouldn’t feel comfortable asking a Be My Eyes volunteer because that’s confidential stuff, and I wouldn’t want to wake up my family either. I’d use Be My AI or Seeing AI. I don’t trust AI more than my family, but I do trust it more than volunteers (P13)''}&	Privacy, Trust	& Family\\
Rental document&	 \textit{``I have used Seeing AI for my rental doc so many times, short text feature is quite great. I wanted to make sure I get the all the main details.''} (P19)& Convenience	& Family\\
Bill \& receipts&	 \textit{``I checked a bill with seeing AI currency feature several times with the wrong result and ended up in having help from her daughter (P2)''}&	Exploratory& Family\\
Credit cards&	\textit{"I check credit card in my junk mail whether it’s a credit card offer or something from my bank or other places (P10)"}&	Exploratory &	Aira\\ 
Debit card	&\textit{``I'd gotten a new debit card and needed to make sure I was throwing away the old one. I used it to read the card numbers and confirm which was which. - (P19)''}	&Convenience & Family\\
Handwritten check	& \textit{``I’ve used Be My Eyes AI to read a handwritten check in a pinch, but I knew there was a risk—it could’ve exposed bank info (P9)''} & Convenience & Family\\
Utility bills&	 \textit{``I don't want to share every detail with my family, so I use Seeing AI and sometimes ChatGPT for a bit of testing''} (P9) &	Privacy & Family\\
Rent statement&	\textit{``I wouldn’t feel comfortable asking a Be My Eyes volunteer because that’s confidential stuff.''}(P8) &Privacy & Family\\
Insurance&	\textit{``If it’s something I need to read, especially if it’s not in braille, my employer’s insurance, not comfortable to call a volunteer.''} (P10)&	Privacy & Family\\
Currency &	 \textit{``But sometimes the Microsoft AI currency reader gets it wrong. I’ve had it say `15 10,' and I end up double-checking with my daughter.''} (P15)&	Privacy & Family\\
\hline

\textbf{Digitized Physical Media}&	\\	
Official email& \textit{``I do share a lot of theoretically private information with ChatGPT, like writing letters or translating emails for official purposes.''} (P12)&	Exploratory&family, screenreader\\
Unopened junk mail & 	 \textit{``I’ve used it to read some mail too, or at least to get the general idea of what the mail is about.''} (P4)&	Exploratory &aira\\
Official flyer &	\textit{``I frequently use Seeing AI, ChatGPT, Be my AI, JAWS AI. Gemini for document interpretation and understanding tasks, such as flyers or product instructions''} (P3)&	Exploratory& aira, family\\
 Email &	\textit{``Nowadays I use ChatGPT a lot for this''} (P15)&	Exploratory&themselves\\
Websites &	\textit{``websites, like the Health Equity. Even printing pages can be tricky. For quick checks, I use Be My Eyes or Seeing AI, depending on which gives the better description.''} (P14)&	Exploratory &family, screenreader\\
ID- Verification&	\textit{``I’ll ask if I can bring it in person, share a download link, or set up a Zoom call and hold up my ID. Then I also use now Seeing AI to scan the page to tell me if it looks right''} (P11) &Privacy&Aira, screenreader\\

    \bottomrule
    \end{tabular}
    \caption{Current Use of GenAI for Visual Privacy with Representative Quote, Perceived Benefit, Pre-GenAI Alternatives}\label{tab:current-practice}
\end{table*}

\section{RQ1: Current Use of Generative AI Tools for Visual Privacy}
In this section, we present several themes that illustrate how people with visual impairment engage with GenAI tools to manage their visual privacy across diverse personal, social, and professional contexts (Table~\ref{tab:current-practice}). We also present nuanced privacy judgments and risk assessments users undertake when choosing to rely on such technologies.

\subsection{Environmental Awareness and Spatial Privacy in Private \& Public Spaces with GenAI}
Participants frequently mentioned their usage of GenAI tools to scan their surroundings in both private and public settings. 
For instance, P15 explained a common case: \textit{``Say, I had to take a photo of myself. Then I want to know that there's not a bra hanging down behind me, or something. I don't want to give a weird impression when posting my picture. Before GenAI, I would often send it to my mom or friend to check for me.''} This illustrates how participants consider social or reputational risks of unintended background content and cases where they replaced human assistance with GenAI. 

Some participants also discussed GenAI in supporting navigation, such as airports or train stations, to read sign or unfamiliar spaces. P21 noted, \textit{``Earlier I used to ask someone and show my phone on the map. It's a bit weird to give my phone to someone to ask for directions. I hope they are not clicking on anything else.''} This reflects the privacy-aware calculation of interpersonal and device sharing risks, including inadvertent data exposure and device misuse. 
A few participants described using GenAI to locate gendered facilities, such as restrooms, to avoid potentially awkward or privacy-compromising interactions. These scenarios illustrate how blind and low vision people actively assess situational risks and use GenAI tools to navigate both functional needs and privacy considerations.

\subsection{Navigating Privacy in Digital Social Life with GenAI}
A common use of GenAI among participants was to inspect image backgrounds before sharing photos on digital platforms. Participants described using Seeing AI and Be My AI to check for unexpected or sensitive objects, visible personal documents, general clutter, or the presence of children. 

Some participants shared broader concerns about the growing autonomy of GenAI tools. P15 explained how she use Seeing AI, Be my AI, and chatGPT to frequently get image descriptions \textit{``I have the apps describe a photo, and I was in a dinner and I don't share images of my child in social. I asked if there’s a child in the photo before.''} P12 shared his experience of using GenAI for both image generation and privacy management-\textit{``I asked ChatGPT to turn a photo into an anime-style image of riding a rocket with sunglasses on, it was for a fun birthday post for my friend. Original photo was taken in front of my apartment. I used Be my AI to make sure my apt number was not visible.''}. This highlights concerns about embedded metadata or identifiable background elements in generated images. 

Similarly, P6 shared his practice for not posting certain images on social media and a similar practice in avoiding GenAI from processing images taken in certain cases. He explained \textit{``If I take a photo at a high-end restaurant on my friends' birthday, I would not process that info in GenAI since I think I am being surveilled, similarly, I would not post it on social media. I could lose my disability benefits. I am giving AI no opportunity to take my disability benefits away from me.''} This highlights a risk-aware strategy, participants navigate in response to institutional gate-keeping as well as awareness of algorithmic inference, which can invalidate one’s eligibility for public support. 


Similarly, some participants mentioned using character-based GenAI tools like Kindred or Replika and character.AI as emotional companions. As P17 said \textit{``I use Replika with a voice I like as a private space to share personal thoughts without fear of judgment or surveillance. You know many thing I can not always do with family.''} She also expressed uncertainty about potential risks, particularly whether the app retained memory of their conversations. This underscores a broader definition of privacy encompassing affective and relational dimensions where GenAI serves as a trusted confidant.


\subsection{Practices of GenAI for Sensitive Content}
Participants described varying levels of comfort and caution when using generative AI.tools to interpret documents and physical objects containing sensitive information (Table~\ref{tab:current-practice}). 

\textbf{Medical Content.} Participants expressed caution when dealing with medical content while mentioning \textit{``privacy''} as a reason to prefer GenAI tools over traditional practices. P16 mentioned \textit{``I avoid GenAI tools for sensitive documents like \textit{medical records}. I still go for trusted humans (e.g., family or Aira agents under NDA agreements).''} In contrast, P17 mentioned using Generative AI for interpreting medical records, such as mammograms and MRIs, for better explainability and privacy,
\textit{``
I would call my sister and tell her to ask my brother-in-law, or make another appointment. I used Be My AI and ChatGPT to make sense of my report. Honestly, I liked the idea of being private.''} 
Some participants used GenAI tools to read sensitive physical items. P1 shared: \textit{``I used AI for my pregnancy test. It said positive several times, and I was stressed. I took more pictures, and then it said negative. I was relieved.''} This highlights how participants preferred using AI independently, even when results varied, to maintain emotional and informational privacy in personal situations.

\textbf{Financial Document.} Participants reported adjusting their methods based on perceived sensitivity and situational privacy needs.
P9 said \textit{``For my credit card, I’d rather use a bonded volunteer. but I have used Be My Eyes AI to read a handwritten check in a pinch, but I knew there was a risk—it could have exposed bank info. Sometimes I just need to know what is in the memo line [..] I don’t want to bug anyone. I try to use secure options like Aira’s free call, which is limited.''} 
 P19 on the other hand, used Seeing AI to read text on a debit card: \textit{``One time I used Seeing AI. I needed to make sure I was throwing away the old debit or new. I used it to read the card numbers and confirm which was which.''} Some noted that their choice depends on the level of trust and availability.

\textbf{Digitized Physical Media.}
Participants also used GenAI tools to navigate digitized and potentially sensitive physical content. 
P11 expressed caution with document uploads, even on trusted platforms: \textit{``am careful with uploading documents/photos, particularly Health Equity one. Even with Aira, who has confidentiality agreements, I am picky. I'd rather call my mom or dad. I sometimes use Seeing AI as a last resort.''}
P20 reflected on using Microsoft Seeing AI to sort mail \textit{``Seeing AI helps me sort mail. There might be something very private in that letter, and I wouldn’t even know when I opened it with AI.''} This reflects a privacy pragmatism mindset where participants recognize the potential sensitivity of what they are sharing but justify the decision based on context. 

\subsection{Sociopolitical Decision Making Factors in Adopting GenAI Tools}

We observed some decision-making factors of adopting GenAI, which is shaped by broader concerns about access, policy assumptions, and representational harms. 

\textbf{Affordability of GenAI.} Participants raised concerns about the cost of GenAI technologies and the assumptions behind their pricing. P11 echoed these concerns, criticizing the pricing and accessibility of tools like Envision AI and noting disappointment in how products are marketed and supported.
P11 shared 
\textit{``Tools like Envision AI are costly and not fully accessible. Meta’s smart glasses offer good scene descriptions, but their partnership with Be My Eyes over Aira felt like privacy wasn’t a priority. I’d like to see better privacy controls before I fully rely on it.''}

\textbf{AI policy decisions around race and people descriptions can lead to incorrect decision-making.} A few participants reflected on how overly cautious content filters can reduce the utility of AI-generated descriptions. As P18 noted: \textit{``As a blind person, I rely on AI for details that sighted people often take for granted. But sometimes the descriptions are so cautious about avoiding potentially sensitive language that they leave out useful context. For example, if I am posting a photo and it just says ‘two people’ without identifying who is who like ‘Jennifer is on the left and Rita’s on the right’ or omits visual markers like skin tone or gender, the information becomes less helpful. I understand the privacy concerns, but overly vague descriptions limit the usefulness of these tools.''}

\textbf{Myths of AI on Disability.} Some discussed the complexities of using generative AI tools as a blind user, particularly when it comes to disability representation and data sharing. P11 also shared firsthand experiences with AI tools repeating outdated stereotypes: \textit{``I asked ChatGPT what it knew about blind people, and it responded with myths like ‘they have stronger other senses’ or ‘they need to be taken care of.’ It didn’t acknowledge the mistake even when I corrected it. Another time, we asked about Braille dots for a letter, and it gave three different, incorrect answers.''} 

\section{RQ2: Perceived Limitations and Expectation of GenAI in Managing Privacy}
In this section, we present participants' contextual privacy norms across several common scenarios (explained in the method section~\ref{Methods_protocol}). We highlight their perceived limitations and challenges of using GenAI to manage privacy in these contexts, and how current experiences and perception shaped their expectations for GenAI, ultimately informing future design recommendations (Section~\ref{design}). 

\vspace{-2mm}
\subsection{Scenario 1: Self Appearance: Impression Management} 
\vspace{-2mm}
While the majority didn't use GenAI for this purpose, participants shared both skepticism and cautious optimism regarding the use of GenAI tools for self-appearance management. As practice, some relied on touch and internal cues to assess their appearance, while some called friends or family for feedback, as well as Be My AI or Aira to check for issues like stains on clothing. P11 illustrates how GenAI would be useful~\textit{``It’s kind of uncomfortable to think about, I remember a specific time Aira saved me, my bra was showing. I was thinking of using Seeing AI.''} P11 emphasized that she would consider GenAI if the tool guaranteed complete privacy, supported local face processing, and could simulate empathy or understanding, especially in tasks involving bodily presentation or vulnerability. 

\underline{Limitation \& Expectation.} We noticed varied expectations while some emphasized the importance of GenAI tools being transparent about their capabilities the extent to which they can provide support, for instance, P18 said {\textit{``Does it just say a wrinkle or spot on dress, or exactly where it is like a tactile like feedback and with a level of certainty?''}}. P5 viewed visual self-presentation with more importance \textit{``I wouldn’t be wearing anything provocative. It wouldn’t be like a `baby do me look. I would prepare based on occasions like date, funeral, movie, so I wouldn't have concerns about people seeing me and judging.''} 
Overall, participants expressed a spectrum of expectations for GenAI tools, including privacy, bodily presentation. 

\subsection{Scenario 2: Indoor Spatial Privacy} 
We noticed two groups among participants: (a) those who have begun integrating these GenAI tools for indoor spatial privacy in addition to traditional methods, and (b) those who continue to rely on tactile methods taught in blindness rehabilitation programs. 
Across both groups, participants commonly relied on touch to assess and manage the presentation and cleanliness of their spaces. 

Participants often described managing indoor spatial privacy as important not only for personal safety but also for ensuring that their living or working spaces were socially acceptable from the perspective of others.
Some participants acknowledged that GenAI could potentially streamline and augment these checks, for instance, P3 said, \textit{``Now that I am really thinking about it, instead of calling someone, using AI with the video would make it go faster. I would imagine GenAI integration to video streaming platforms to check clutter, and particularly background checks when I am in a video call.''} 

\underline{Limitation \& Expectation.} Participants valued hands-on spatial checks, but emphasized augmenting those checks with GenAI rather than replacing them. As P13 expected, GenAI to use camera input and then align with tactile and haptic-based cues {``There is clutter of used clothes two feet away. Then provide some vibration to indicate that.''} 
Some shared their expectation of using GenAI when ``private mode'' is established to limit AI sensing in sensitive categories. P6 said \textit{``I have a lot of things lying around, sometimes some private magazine I don't want to be captured during environmental scanning.  My question would be, is this happening locally and  without cloud upload?''}

\subsection{Scenario 3: Sharing Visual Content in Social Media}
Participants considered GenAI as useful for quick facts and clarifications, but some participants preferred human assistance for socially sensitive or subjective decisions. 

\underline{Limitation \& Expectation.} Participants also discussed their expectations of GenAI which include seeking clarity on background detection, descriptive querying, identity cues, and human-in-the-loop options in this scenario of sharing visual content in social media. 
Some also highlighted challenges in shared living spaces, where visual ambiguity arises due to overlapping personal items. As P10 explained: \textit{``I have a roommate we share space sometimes sensitive things lie around. Might not be mine, but hers. If those are in the picture, I don’t know how to manage. GenAI could help maybe.''} P10 expected a way to distinguish between self vs others' and privacy-aware tagging of visual content in a social media context. P5 expected support for identity matching so she can verify who is in the image without guessing.

\subsection{Scenario 4: Visual Content Privacy when Sharing with Others}
Participants described a range of privacy expectations when using GenAI to access and extract content from documents, especially before sharing with others. They frequently mentioned employers, family/friends, and government. 

\underline{Limitation \& Expectation.} Many participants emphasized the need for secure processing of sensitive documents, such as, ID, credit cards: P14  \textit{``I am worried, but also think it would have been very helpful to use LLM. Some way to not send my document to cloud or just to see `your file is processed with industry-standard encryption will make a huge difference.' ''} 
P9 also mentioned a potential way to set privacy level based on recipient \textit{``I would like to assess privacy level: Review this photo for sharing with family, employer, or HR.''} 

Some participants described strategies to minimize re-identification risks. P11 shared a compartmentalization tactic: \textit{``I am a person who sends username in one email and password in another. I use GenAI same way I upload or capture documents in multiple rounds to make it harder for someone to get all the information if there’s a leak.''} P9 mentioned strategies for routinely deleting content immediately after task completion. There was a shared expectation for more control over data processing and storage, including the ability to assess task-specific risks for highly sensitive content.

\subsection{Scenario 5: Visual Privacy Management by BLV Professionals}
BLV Professionals expressed a tension between the accessibility benefits of GenAI tools and the risks to visual privacy, particularly when handling sensitive documents of others. Several participants working in financial, educational, and social service sectors shared workplace-specific experiences.
Some described practices to protect client confidentiality. P11, a former accessibility tester \textit{``I reviewed documents with proprietary client information. We set up a closed system, and we avoided bringing in third parties. Honestly, I know GenAI will be mainstream. It’s just a matter of time and the highest bidder.''} 
Moreover, institutional professionals like P9 reported segmenting their GenAI use based on data sensitivity- \textit{``I try not to input any student documentation, grade info into Gemini right now, but I use LLMs to create material for training purposes. We are actively discussing how to institutionalize GenAI use.''}
A blind professional, P21, reported using generative AI tools in roles that involved handling information, such as reviewing documents to determine eligibility for benefit programs. 

\underline{Limitation \& Expectation.} Participants emphasized using GenAI tools within secure, enterprise-approved environments (e.g., Office365 accounts with institutional safeguards).
P13 described using GenAI to review instructional documents, noting a lack of privacy concerns due to the nature of the materials: \textit{``The documents I used weren’t mine, they were for a lesson, and they were pretty old, so I wasn’t really concerned about privacy.''} They also recalled trying various AI tools like TapTapSee and Seeing AI for personal tasks, but found the results inconsistent or unhelpful.
P12, who works in the financial sector, highlighted the situational use of GenAI tools and shared a preference for AI-based image recognition apps like Seeing AI over asking others for assistance \textit{``For casual tasks like reading currency or simple documents, Seeing AI is my go-to. It’s fast, reliable, and does exactly what I need. I’d still ask my boss before using AI on confidential work documents—because companies can have their own rules.''}

\subsection{Scenario 6: Outdoor Spatial Privacy }
In this case, many expressed use of GenAI, which is shaped by past training, current usability limitations, and broader social concerns. For instance, P13 shared skepticism toward overreliance on technology. However, she acknowledged growing community interest in tools like GoodMaps~\footnote{https://goodmaps.com/}, especially as their coverage of indoor spaces expands. Others echoed similar tensions. P12 described how GenAI tools such as Seeing AI felt socially awkward in public spaces: \textit{``You have to hold the phone up as the bus approaches... people can think you’re taking their photos.''} 

\underline{Limitation \& Expectation.} Many shared expectations towards ergonomic, hands-free, and privacy-aware design. P11 described the unsolicited and time-consuming experience of being placed in a wheelchair at airports without understanding the range of blindness \textit{``I would really prefer privacy-preserving GenAI that guides without requiring public disclosures and unsolicited wheelchair placement in airports. I can see an option with GenAI to support step-by-step instructions or a human hand-off?''} P9 shared concerns about location privacy \textit{``I don't want people to know where I am travelling. I have envisioned and meta glass, but I am not sure if they process GPS and camera data locally or if they connect to other apps externally without my consent.''}

\section{Discussion}
We presented one of the first investigations into current practices and design opportunities of GenAI for visual privacy among people with visual impairments. Below, we synthesize our findings to show how GenAI-mediated visual privacy is shaped by personal agency, social norms, and technological constraints.

\subsection{Notion of Visual Privacy within GenAI}
Our study surfaced two interrelated dimensions of visual privacy for people with visual impairments when interacting with generative AI tools: (1) impression management on how individuals control the way others perceive their appearance and environment, and (2) accountability in handling others’ private visual content. 
 
 Our participants expressed mixed feelings about relying on GenAI, similar to prior research on crowd-powered accessibility tools like VizWiz~\cite{bigham2010vizwiz}. 
 While our findings indicate the convenience of the GenAI tool in addressing long-standing challenges in managing one's privacy independently, the impersonal nature and lack of contextual awareness of GenAI could potentially amplify privacy concerns. 
 Participants often use GenAI for socially sensitive tasks that directly shape how others perceive them and their surroundings, or convey specific messages (e.g., competence, professionalism), even for curating the visual backdrop during video calls. These tasks carry implications for social signaling from Goffman’s theory of impression management~\cite{tseelon1992presented}. However, the generic, decontextualized feedback from GenAI systems often failed to meet users’ nuanced expectations. For instance, our findings highlight that privacy management varies across different scenarios, while in one case, risks are centered on the user’s own visual contents, in another, risks may arise from shared environments or unintended bystanders. These distinctions underscore the importance of context and relational expectations (e.g., Nissenbaum’s contextual integrity~\cite{nissenbaum2004privacy}). 
 
Moreover, our findings point to systemic ambiguities in the institutional handling of GenAI use. Participants, particularly blind professionals, described being left to self-govern the use of GenAI tools for high-stakes tasks such as processing confidential disability paperwork, verifying identification, or interpreting student documents. Despite recognizing both the benefits and risks, the lack of formal guidance, secure infrastructure, or vetted tools leads users to make complex privacy decisions independently. This reflects what Ahmed et al.~\cite{ahmed2016addressing} and Akter et al.~\cite{akter2022shared} have identified as structural vulnerabilities, where disabled users shoulder disproportionate responsibility in mediating privacy risks in the absence of institutional safeguards.

\subsection{Design Gaps: Privacy-Preserving GenAI Ecosystems}
While prior research has examined visual privacy for BLV individuals, particularly through human-mediated services like Aira and Be My Eyes~\cite{stangl2020visual, akter2020uncomfortable}, our study extends this work by shedding light on how BLV users re-purpose GenAI tools for both accessibility and privacy-sensitive tasks. While existing research has explored how blind individuals use visual interpretation tools like Aira, Be My Eyes, and Seeing AI, and some emerging tools~\cite{adnin2024look}, we highlighted the design gap of GenAI tools for visual privacy management. 

\revise{Our findings indicated GenAI’s distinct affordances to emphasize the potential for context-aware, customizable, and conversational forms of privacy mediation, in contrast to rule-based or static vision systems.} 
For example, participants preferred GenAI over human assistance for sensitive tasks like reading pregnancy tests or mammograms, yet also expressed concern over data storage~\cite{sharma2024m} and potential model training risks~\cite{zhou2024rescriber}. These practices highlight that visual privacy involves not just what is shared, but how and with whom, introducing interpersonal and emotional dimensions of risk.
In managing spatial privacy, participants emphasized complementing, rather than replacing, tactile methods with GenAI. This underscores a hybrid design space not yet central in GenAI discussions. Similarly, our findings add to research on online sharing and impression management~\cite{Park20, Bennett18} by surfacing the challenge of distinguishing between one’s own and others’ belongings in shared environment an area yet to be integrated in mainstream GenAI tools. 
Unlike rule-based tools, which operate on fixed detection pipelines, GenAI allows users to dynamically shape and refine their queries, thus embedding privacy preferences into interaction flows. 

We suggest GenAI design should move beyond basic image description toward adaptive, context-aware, and multimodal feedback that reflects users’ privacy priorities. This includes features like personalized object filters and proactive alerts for sensitive content grounded in the real practices and concerns of BLV users.

\subsection{Sociopolitical Barriers, Policy Implications and Risk of GenAI Censorship}
\revise{
Our findings highlighted that the interaction of BLV users with GenAI tools is influenced not just by technical capabilities and personal preferences, but also by sociopolitical structures. Several participants shared concerns related to the absence of explicit institutional safeguards when employing GenAI tools in situations that are seemingly in need of legal protections, such as the affordability of technology integrated with GenAI. While the Assistive Technology Act of 2004 mandates that states ensure access to assistive technologies~\cite{bausch2005assistive}, participants shared how the high costs of advanced AI-based tools (like Envision Glasses) place an unfair burden on users.

Our findings also shed light on the restricted GenAI functionality, which may limit BLV users' ability to access and share content. For example, over-filtering or censoring essential indicators related to appearance, gender, or race, likely influenced by platforms’ bias mitigation policies, lead to vague outputs which undermine both accessibility and utility. Under the Americans with Disabilities Act (ADA)~\cite{ada1990americans}, individuals with disabilities should have the same opportunities and access as non-disabled individuals, legislation that has been increasingly applied to digital environments, including AI tools~\cite{neogi2024protecting}. However, our findings underscore that blanket policies (e.g., redacted descriptors, broad safety filters) may potentially conflict with BLV users’ needs for detailed, descriptive, and context-rich feedback.

Participants also described reliance on GenAI in understanding sensitive documents (e.g., insurance forms, student records). In professional settings, this situates BLV individuals at the intersection of ADA compliance and institutional data protection mandates such as, HIPAA, FERPA. While there are guidelines, such as, FCC’s CVAA (2010)~\cite{nopublic} for accessible communications and ADA Title I~\cite{adaEmploymentTitle} for workplace equity, participants are at a crossroads without institutional policy guidance. We believe addressing these tensions needs more than guidelines; rather, fundamental rethinking of GenAI design for an enterprise-safe and privacy-preserving ecosystem that both complies with data protection while allowing BLV professionals to leverage the benefits of GenAI meaningfully. 
}

\subsection{Design implications}
\label{design}
Drawing upon BLV people's perceived limitations and expectations of GenAI, we suggest actionable design implications.


\textbf{On-Device or Local Encryption.} Our findings revealed a common expectation across all scenarios for local, on-device data processing,  particularly for facial images, home environments, financial or medical documents, and GPS or camera inputs used by BLV individuals in both public and private spaces. This expectation also extended to cases where BLV users processed documents on behalf of others in professional roles (e.g., student papers, social security benefit documents for eligibility). 
One potential approach is to design a modular, plug-and-play GenAI model sandbox per data type: Visual, Text, Audio/Video, and Sensor Module (GPS) isolated in a trusted execution environment (e.g., Intel TDX, iOS Secure Enclave) ~\footnote{https://docs.trustauthority.intel.com/main/articles/concept-tees-overview.html}). This technical design could be paired with audit logs and explainability layers to provide privacy logs detailing accessed data, module use, and outputs. This will ensure there is no network access for the sandbox unless explicitly approved by the user.

\textbf{Federated Compliance-Aware GenAI: Secure Toolkit for BLV Professional.} 
Our findings revealed a distinct concern around protecting others’ privacy when BLV professionals use GenAI in institutional contexts like banking, education, or social services, where users often process sensitive data on behalf of others.
In response, we propose a secure GenAI toolkit that integrates with compliance-aware platforms (e.g., SharePoint, OneDrive, Google Workspace) to handle documents locally or in secure federated and decentralized environments~\cite{massonet2011monitoring, sharma2024can}.  It can also enforce role-based access controls (RBAC) and aligns with institutional data protection standards, such as FERPA and HIPAA, through privacy-policy-aware schema. To further safeguard privacy, the system includes post-task accountability prompts using named entity recognition (NER) to flag sensitive data (e.g., names, birthdates) and guide protective actions.

\textbf{Personalized, Privacy-Aware Appearance Feedback System.} Our findings show interest of using of GenAI for self-appearance management, yet this raises privacy concerns and the need to distinguish bodily privacy (e.g., exposed undergarments) from aesthetic issues (e.g., smudged makeup). To support BLV users, we propose integrating appearance-specific descriptors through a fine-tuned multi-label visual classifier and customizable sensitivity profiles. the system can also feature memory-based look validation, which compares the user’s current appearance with previously approved outfits using CLIP-based~\cite{peng2006clip} visual similarity and securely stored, user-labeled photos.



\textbf{Visual Disambiguation for GenAI for Shared Space Privacy.}
To support blind users in managing visual privacy in shared environments, GenAI tools should incorporate personalized object recognition and context-aware disambiguation. A technical solution involves allowing users to train the system to identify and differentiate their personal items from those of others using labeled image examples. For instance, when analyzing images, the system can prompt: \textit{``Multiple items detected, do you want to identify which are yours?''} and apply role-specific visual tagging (e.g., not yours, shared, uncertain).

\subsection{Limitations}
\revise{Our work has a number of limitations. Our participants' sample of 21 may not be fully representative of the broader BLV community. From our study, very few participants had the affordability to use emerging assistive tools, such as Meta Rayban Glass, Envision Glass, etc, for some use case relevant to visual privacy, which might suggest limited representation of higher socio-economic groups. 
}

\section{Conclusion} 
Our paper presents one of the first in-depth empirical studies on how blind and low vision (BLV) individuals use Generative AI (GenAI) tools to manage visual privacy across diverse everyday contexts. Our findings extend the existing literature on both privacy and accessibility by demonstrating that visual privacy for BLV users is not just about limiting data disclosure, but also about facilitating autonomy, trust, and social accountability. As GenAI continues to evolve, we call on researchers and designers to prioritize the lived realities of marginalized users not as edge cases, but as starting points for innovation.

\begin{acks}
 \revise{We thank our participants for their thoughtful insights. This work was partially supported by the National Science Foundation SaTC grants (\#2125925, \#2148080, \#2126314). 
 
Conflict of interest disclosure: Leah Findlater is also employed by and has a conflict of interest with Apple Inc. Any views, opinions, findings, and conclusions or recommendations expressed in this material are those of the author(s) and should not be interpreted as reflecting the views, policies, or position, either expressed or implied, of Apple Inc.

 }
\end{acks}


\bibliographystyle{ACM-Reference-Format}
\bibliography{sample-base, references}

\appendix

\section{Appendix}
\label{appendix}

For these informal interviews, I’m only considering people who used GAI before.
I am first going to ask you a round of questions so that I can get a better grasp of your current practice and experience with Generative AI Tools. [“Generative AI for Privacy” if any]
\subsection{Interview}
\textbf{Section 1: RQ1: How and for what purpose do blind users use Generative AI tools?}

What generative AI tools are you currently using?
\begin{itemize}
\item ChatGPT
\item DALLE.2
\item Be My AI
\item Envision AI
\item Gemini
\item Claude
\item Others
\end{itemize}

Which ones have you used specifically to access visual information, like images, videos, or real-world visual information?

\begin{itemize}
\item Be My AI
\item Envision AI
\item ChatGPT
\item Others
\end{itemize}

For each of the tools you mentioned, what purposes do you use them for? Let's start with one by one.

\begin{itemize}
\item Can you share an experience or event when you used this tool (e.g., Envision AI)?

\item What makes you decide to use this tool? (Some factors: they might share some insights on its usefulness, performance, new features, etc.)

\item Could you briefly explain how this tool works? Maybe you could provide an example and walk me through your process of using this tool.

\item Great, have you found these features in any existing tools you previously used? What is different about generative AI compared to past tools you used for this similar purpose?

\item It's great to hear about how and for what purpose you use this tool. Is there any other purpose you consider using this tool for? If not, can you think of any other way this tool could be useful in your daily tasks?
\end{itemize}

Based on their response, if no privacy-related use cases for Generative AI tools are mentioned, then proactively ask about privacy-related tasks for each tool they mentioned:
\begin{itemize}
\item Have you used these tools for any purpose related to privacy? (Open-ended for them to answer)
\end{itemize}

\textbf{Definition of the Scope:} Visual Privacy for Blind Users refers to the safeguarding and management of sensitive visual information that could be shared or disclosed through the use of Generative AI tools. This includes but is not limited to the protection of content such as medical records, financial information, or any other visual data that might be considered private when engaging in daily activities. For instance, when using these tools to receive descriptions of potentially sensitive content—whether it’s medical or financial records or when scanning surroundings for navigation.

Next, I will ask you about specific scenarios to better understand how you currently use Generative AI tools and to explore your thoughts on how you might want to use them in similar situations in the future.

\textbf{Section 2: Future Design Preferences for Generative AI Tools}

\textbf{Privacy Scenario 1 (Own Privacy): Checking Self Before Seeing Other People} Imagine you are about to meet someone at a social event or professional meeting. Before you step out, you want to make sure your appearance is in order—clothes are neat, hair is tidy, and nothing is out of place. You decide to use a Generative AI tool to describe your appearance.

\begin{itemize}
\item How would you currently use a Generative AI tool to check your appearance before meeting someone?
\item How would you like this tool to be designed in the future to better assist you in this scenario?
\item What is it about GAI that makes you think this way (risk perception)?
\end{itemize}

\textbf{Privacy Scenario 2 (Own Privacy): Checking House Before Having Someone Over}
Imagine you’re expecting guests at your home and want to ensure that your space is tidy and presentable. You want to check for things like clutter in the living room, any dishes left out in the kitchen, or if there’s anything unusual that might be out of place. You decide to use a Generative AI tool to help you assess your surroundings.

\begin{itemize}
\item How would you currently use a Generative AI tool to check your house before having someone over or in some similar situation?
\item How would you like this tool to be designed in the future to better assist you in this scenario?
\end{itemize}

\textbf{Privacy Scenario 3 (Own Privacy): Picture to Share on Social Media}
Imagine you’ve taken a photo and are considering sharing it on social media. Before posting, you want to ensure that the image doesn’t contain any private information, such as personal things, recognizable locations, or other details that you wouldn’t want to be publicly visible. You decide to use a Generative AI tool to analyze and describe its contents.

\begin{itemize}
\item How would you currently use a Generative AI tool to check the content of a picture before sharing it on social media?
\item How would you like this tool to be designed in the future to better assist you in this scenario?
\end{itemize}

\textbf{Privacy Scenario 4 (Own Privacy): Document Picture to Share With Employer}
Imagine you’re preparing to share a financial report with an employee, but before doing so, you want to ensure that the document doesn’t contain personal salary details, confidential company financials, or other data that you don’t want to disclose.

\begin{itemize}
\item How would you currently use a Generative AI tool to check such a document before sharing it with an employee?
\item How would you like this tool to be designed in the future to better assist you in this scenario?
\end{itemize}

\textbf{Privacy Scenario 5 (Others' Privacy): Reading Private Documents (Social Security) for Issuing Benefits for Employees} As an employer, you are responsible for issuing benefits to your employees, which requires you to review private documents, such as Social Security information, tax forms, or other sensitive personal data. You decide to rely on a Generative AI tool to assist you in reading this information while ensuring that privacy and confidentiality are maintained.

\begin{itemize}
\item How would you currently use a Generative AI tool to read and process documents of your employees in such a scenario for issuing employee benefits?
\item How would you like this tool to be designed in the future to better assist you in this scenario?
\end{itemize}

\textbf{Privacy Scenario 6 (Others' Privacy): Scanning Surroundings When Outside (Navigating the Airport)}
Imagine you are at an airport and need to navigate through various areas, such as finding your gate, locating restrooms, or identifying nearby amenities. You decide to rely on a Generative AI tool to help you understand and navigate the environment.

\begin{itemize}
\item How would you currently use a Generative AI tool to scan and understand your surroundings in an airport?
\item How would you like this tool to be designed in the future to better assist you in this scenario?
\end{itemize}

\end{document}